\documentclass[aps,pra,graphicx,9pt,twocolumn,superscriptaddress]{revtex4-2}
\usepackage[colorlinks,linkcolor=blue,urlcolor=blue,anchorcolor=blue,citecolor=blue]{hyperref}
\usepackage{amsmath}
\usepackage{graphicx}
\usepackage{dcolumn}
\usepackage{mathrsfs}
\usepackage{amssymb}
\usepackage{amsfonts,multirow}
\usepackage{bm,changes}
\usepackage{setspace}

\begin{document}
\title{Geometric criticality in the driven Jaynes-Cummings model}

\author{Ken Chen}
\author{Jia-Hao L\"{u}}
\author{Hao-Long Zhang}
\author{Fan Wu}
\author{Wen Ning} \email{ningw@fzu.edu.cn}
\address{Fujian Key Laboratory of Quantum Information and Quantum\\
Optics, College of Physics and Information Engineering, Fuzhou University,
Fuzhou, Fujian 350108, China}

\author{Zhen-Biao Yang} \email{zbyang@fzu.edu.cn}
\author{Shi-Biao Zheng} \email{t96034@fzu.edu.cn}
\address{Fujian Key Laboratory of Quantum Information and Quantum\\
Optics, College of Physics and Information Engineering, Fuzhou University,
Fuzhou, Fujian 350108, China}
\address{Hefei National Laboratory, Hefei 230088, China}

\begin{abstract}
When the photonic mode in the Jaynes-Cummings model is driven by an external
classical field, the system can undergo the photon-blockade breakdown phase
transition at a critical point. Such a phase transition has been detailedly 
investigated, but the critical properties of the eigenstates remain largely
unexplored so far. We here study the geometric criticality associated with
these eigenstates. The amplitude and phase of the drive serve as the
control parameter of the governing Hamiltonian. We find the quantum metric
and Berry curvature tensors for each eigenstate display divergent behaviors
in the critical region. More importantly, the divergence associated with
bright eigenstates is much more pronounced than that for the unique dark
state. Our theoretical results can be experimentally confirmed in circuit
quantum electrodynamics systems, where the driven Jaynes-Cummings model has
been realized.
\end{abstract}

\maketitle  

\section{INTRODUCTION}
In recent years, increasing interest has been paid to quantum
geometry \cite{Toermae2022Superconductivity,peotta2025Quantum,Toermae2023Essay,Liu2024Quantum}, quantified by the quantum geometric tensor (QGT) \cite{Provost1980Riemannian}, which
describes the inherent geometry of the eigenstate space of the Hamiltonian
governing the behaviors of a quantum state. The real part of the QGT, which
is symmetric and referred to as quantum metric, measures the distance
between nearby eigenstates of the Hamiltonian. The quantum metric plays a
critical role in understanding important physical phenomena in solid-state
physics, exemplified by flat-band superconductivity \cite{Liu2024Quantum}. Its imaginary part,
which is antisymmetric and known as Berry curvature \cite{Berry1984Quantal}, provides
information about phase change of the eigenstates \cite{Toermae2023Essay}. The integral of the
Berry curvature over a closed manifold yields the Chern number, which
quantifies the topology of the manifold. The Berry curvature can be thought
as the fictitious magnetic field emanated by a topological monopole, and is
responsible for intriguing electronic transport phenomena, e.g., quantum
anomalous Hall effect \cite{Yu2010Quantized,Chang2013Experimental,Checkelsky2014Trajectory}. In recent years, the QGT has been measured with
different types of pseudospins, including NV centers in diamond \cite{Yu2020Experimental},
ultracold atoms \cite{Chen2022synthetic}, and superconducting qubits \cite{Roushan2014Observation,PhysRevLett.122.210401}, polariton systems
\cite{Flaeschner2016Experimental}, and Bloch electrons in solids \cite{Gianfrate2020Measurement}.

Apart from laying the foundation for topological physics, the QGT provides a
new insight to quantum phase transitions without resorting to order
parameters \cite{Kim2025Direct,PhysRevLett.127.107402,PhysRevLett.99.100603,PhysRevLett.99.095701,PhysRevA.82.012321}. In the proximity of a quantum phase transition, the
eigenstates of the Hamiltonian undergo a dramatic change even if the
control parameter is only slightly changed. As such, it is expected that the
QGT associated with each eigenstate exhibits divergent behaviors near the
critical point. Such critical behaviors have been theoretically investigated
in different systems, including the parametrically-driven nonlinear
resonator \cite{Zhang2024Criticala}, Dicke model \cite{Zhu2024Quantum}, and parametrically-driven Tavis-Cummings
model \cite{Lue2023Quantum}. These systems bear a common feature that the Hamiltonian is
symmetric under the parity transformation. The corresponding phase
transition originates from the spontaneous parity breaking of the parity.
Despite these theoretical advancements, experimental observations of the
critical geometric phenomena are still lacking.

In this paper, we study the critical behaviors of entangled eigenstates of
the driven Jaynes-Cummings model (JCM) \cite{PhysRevA.45.5135,PhysRevX.5.031028,PhysRevResearch.3.023062,PhysRevX.7.011012,Lue2026Critical}, where photonic mode is
coupled to a qubit and driven by an external classical field. This drive
breaks the U(1) symmetry of the Jaynes-Cummings model. Neither does the
driven Jaynes-Cummings model have the parity symmetry. Such a system
possesses a unique dark eigenstate and infinitely many bright doublets. The
quasienergy splittings of the doublets vanish after crossing the critical
point, where the drive strength is equal to half of the coupling strength
between the qubit and the photonic mode. The critical behavior of the qubit
in the dark state has been exploited for realizing robust quantum metrology
\cite{Lue2026Critical}. However, the critical geometric features have not been touched upon so
far. We here investigate quantum geometry for such a critical system. The
calculated quantum metric and Berry curvature for each eigenstate diverge at
the critical point. The critical features associated with bright eigenstates
are much stronger than those for the dark state. As neither the
thermodynamical limit nor the scaling limit is required to observe the
critical behaviors of the driven JCM, this is in distinct contrast with
other qubit-photon systems \cite{Zhu2024Quantum,Lambert2004Entanglement,Ashhab2013Superradiance,PhysRevLett.115.180404}. As such, our theoretical results can
be experimentally demonstrated with presently available techniques, which
would shed new light on critical phenomena in fully quantum-mechanical
light-matter systems.

\section{eigenstates and eigenenergies of the driven JCM}
We begin by considering the driven JCM \cite{PhysRevA.45.5135, PhysRevA.49.1993, feng_exploring_2015}, 
which describes a two-level system resonantly coupled to a quantum field mode while simultaneously driven by an on-resonance signal field.
The interaction Hamiltonian in the interaction picture is given by ($\hbar = 1$ is set)
\begin{equation}
    \begin{aligned}
        H={}&\Omega [ a^{\dagger }| g\rangle \langle e| +a| e\rangle \langle g| +\eta(a^\dagger e^{-i\phi} + ae^{i\phi})/2],
    \end{aligned}
\end{equation} 
where $|g\rangle$ and $|e\rangle$ are the ground and excited
state of the two-level system, 
$a^\dagger$ and $a$ are the creation and annihilation operators for the quantized field mode, 
$\Omega$ represents the coupling strength, 
and $\eta$ characterizes the rescaled amplitude of the driven field with the phase $\phi$.

Next, we solve the eigenvalue problem with
\begin{equation}
    \begin{aligned}
        H|\psi_E\rangle=E|\psi_E\rangle.
    \end{aligned}
\end{equation} 
Here we expand the eigenstate as
\begin{equation}
    \begin{aligned}
        |\psi_E\rangle={}&|\psi_E^+\rangle |e\rangle+|\psi_E^-\rangle|g\rangle,
    \end{aligned}
\end{equation}
where $|\psi_E^+\rangle$ and $|\psi_E^-\rangle$ are field states normalized, satisfying the condition of $\langle \psi_E|\psi_E\rangle=1$.
Therefore, the eigenvalue problem can be expressed as
\begin{subequations}\label{eq:eigenvalue problem}
    \begin{align}
        \Big[\frac{\Omega \eta}{2}(a^\dagger e^{-i\phi} + ae^{i\phi})-E\Big]|\psi_E^+\rangle + \Omega a|\psi_E^-\rangle ={}&0, \label{eq:eigenvalue problem a}\\
        \Omega a^\dagger |\psi_E^+\rangle + \Big[\frac{\Omega \eta}{2}(a^\dagger e^{-i\phi} + ae^{i\phi})-E\Big]|\psi_E^-\rangle={}&0. \label{eq:eigenvalue problem b}
    \end{align}
\end{subequations}

Here we multiply Eq. (\ref{eq:eigenvalue problem a}) and Eq. (\ref{eq:eigenvalue problem b}) on the left by $a^\dagger$ and $a$, respectively, 
which gives rise to
\begin{subequations}\label{eq:eigenvalue problem substitute}
    \begin{align}
        \left\{-\left[\frac{\eta}{2}(a^\dagger e^{-i\phi} + ae^{i\phi})-\frac{E}{\Omega}\right]^2 + a^\dagger a \right\}|\psi_E^-\rangle {}& \nonumber\\
        - \frac{\eta}{2} e^{i\phi}|\psi_E^+\rangle={}&0,  \label{eq:eigenvalue problem substitute a}\\
        \left\{-\left[\frac{\eta}{2}(a^\dagger e^{-i\phi} + ae^{i\phi})-\frac{E}{\Omega}\right]^2 +aa^\dagger \right\}|\psi_E^+\rangle {}& \nonumber\\
        + \frac{\eta}{2} e^{-i\phi}|\psi_E^-\rangle={}&0.  \label{eq:eigenvalue problem substitute b}
    \end{align}
\end{subequations}
By combining Eq. (\ref{eq:eigenvalue problem substitute a}) and Eq. (\ref{eq:eigenvalue problem substitute b}), we obtain
\begin{equation}
    \begin{aligned}
        \mathcal{A}_p(E)\mathcal{A}_m(E)|\psi_E^-\rangle={}&0,
    \end{aligned}
\end{equation} 
where
\begin{subequations}\label{eq:Op Om}
    \begin{align}
        \mathcal{A}_p(E)={}&\mathcal{A}(E)+\frac{1}{2}\sqrt{1-\eta^2}, \label{eq:Op Om a}\\
        \mathcal{A}_m(E)={}&\mathcal{A}(E)-\frac{1}{2}\sqrt{1-\eta^2}, \label{eq:Op Om b}\\
        \mathcal{A}(E)={}&-\left[\frac{\eta}{2}(a^\dagger e^{-i\phi} + ae^{i\phi})-\frac{E}{\Omega}\right]^2+ \frac{a^\dagger a+aa^\dagger}{2},
    \end{align}
\end{subequations}
where $\mathcal{A}_p(E)$ and $\mathcal{A}_m(E)$ satisfy the commutation relation, i.e., $[\mathcal{A}_p(E), \mathcal{A}_m(E)]=0$. 
Therefore, the general solution may be written as
\begin{equation}
    \begin{aligned}
        |\psi_E^-\rangle={}&c_p|\psi_{E,p}^-\rangle+c_m|\psi_{E,m}^-\rangle, \label{eq:psi E}
    \end{aligned}
\end{equation} 
where $|\psi_{E,p}^-\rangle$ and $|\psi_{E,m}^-\rangle$ are solutions to the equations
\begin{subequations}\label{eq:Op Om psi}
    \begin{align}
        \mathcal{A}_p(E)|\psi_{E,p}^-\rangle={}&0, \label{eq:Op Om psi a}\\
        \mathcal{A}_m(E)|\psi_{E,m}^-\rangle={}&0. \label{eq:Op Om psi b}
    \end{align}
\end{subequations}
Using Eq. (\ref{eq:eigenvalue problem substitute a}) and Eqs. (\ref{eq:Op Om}), we can write
\begin{equation}
    \begin{aligned}
        |\psi_E^+\rangle={}&(\frac{\eta}{2} e^{i\phi})^{-1}\{  \mathcal{A}_{p,m}-\frac{1}{2}[1\pm \sqrt{1-\eta^2}] \}|\psi_E^-\rangle, \label{psi E +}\\
    \end{aligned}
\end{equation}
where the $\pm$ notation indicates two alternative forms 
with $+$ corresponding to the p-subscripted form and $-$ to the m-subscripted form.
According to Eqs. (\ref{eq:Op Om psi}) and Eq. (\ref{psi E +}), we have
\begin{equation}
    \begin{aligned}
        |\psi_E^+\rangle=-(\eta e^{i\phi})^{-1} \{ {}& c_p [1 + \sqrt{1-\eta^2}] |\psi_{E,p}^-\rangle \\
        {}&+ c_m [1 - \sqrt{1-\eta^2}] |\psi_{E,m}^-\rangle\}.
    \end{aligned}
\end{equation}

To solve Eqs. (\ref{eq:Op Om psi}), we use squeezing and displacement transformations 
to convert Eqs. (\ref{eq:Op Om psi}) into eigenvalue equations for a harmonic oscillator. 
We multiply Eqs. (\ref{eq:Op Om psi}) on the left by $D^\dagger(\alpha)S^\dagger(\xi)$ and transform the operators $\mathcal{A}_p(E)$ and $\mathcal{A}_m(E)$ using
\begin{subequations}
    \begin{align}
        D^\dagger(\alpha) a D(\alpha)={}&a+\alpha, \\
        S^\dagger(\xi)aS(\xi)={}&a\cosh r-a^\dagger e^{i\theta}\sinh r,
    \end{align}
\end{subequations}
where $D(\alpha)=e^{\alpha a^\dagger - \alpha^* a}$ and $S(\xi)=e^{-\frac{1}{2}(\xi a^{\dagger 2}-\xi^* a^2)}$.
So when we choose 
\begin{subequations}
    \begin{align}
        \alpha={}&-\eta (1-\eta^2)^{-3/4} \frac{E}{\Omega}e^{-i\phi},\\
        \xi={}&re^{-2i\phi}, \quad r=\frac{1}{4}\ln \left( 1-\eta ^2 \right),
    \end{align}
\end{subequations}
Eqs. (\ref{eq:Op Om psi}) can be substituted by the following equations
\begin{subequations}\label{eq:Fock state solutions}
    \begin{align}
        a^\dagger a |\tilde{\psi}_{E;p}^-\rangle={}&\left[(\frac{E}{\Omega})^2(1-\eta^2)^{-3/2} - 1\right] |\tilde{\psi}_{E;p}^-\rangle, \label{eq:Fock state solutions a}\\
        a^\dagger a |\tilde{\psi}_{E;m}^-\rangle={}&(\frac{E}{\Omega})^2(1-\eta^2)^{-3/2}  |\tilde{\psi}_{E;m}^-\rangle, \label{eq:Fock state solutions b}
    \end{align}
\end{subequations}
where $|\tilde{\psi}_{E;p}^-\rangle=D^\dagger(\alpha)S^\dagger(r)|\psi_{E;p}^-\rangle$ and $|\tilde{\psi}_{E;m}^-\rangle=D^\dagger(\alpha)S^\dagger(r)|\psi_{E;m}^-\rangle$. 

The solutions of Eqs. (\ref{eq:Fock state solutions}) are Fock states if and only if 
the constants on the right-hand sides are restricted to the discrete set of non-negative integers.
From Eqs. (\ref{eq:Fock state solutions}), we obtain the discrete eigenenergies as
\begin{subequations}
    \begin{align}
        E_0={}&0,\\
        E_{n,\pm}={}&\pm \sqrt{n}\Omega (1-\eta^2)^{3/4} \quad (n=1,2,...).
    \end{align}
\end{subequations}
The corresponding states are
\begin{subequations}
    \begin{align}
        |\tilde{\psi}_{E;p}^-\rangle={}&|n-1\rangle,  \quad (n=1,2,...)\\
        |\tilde{\psi}_{E;m}^-\rangle={}&\left \{
            \begin{aligned}
                &|0\rangle\\ 
                &|n\rangle,  \quad (n=1,2,...).
            \end{aligned}
            \right.
    \end{align}
\end{subequations} 
Furthermore, we can obtain
\begin{subequations}
    \begin{align}
        |\psi_{E;p}^-\rangle={}&S(r)D(\alpha)|n-1\rangle=|r,\alpha,n-1\rangle,\\
        |\psi_{E;m}^-\rangle={}&\left \{
            \begin{matrix}
                S(r)D(\alpha)|0\rangle=|r,\alpha,0\rangle\\ 
                S(r)D(\alpha)|n\rangle=|r,\alpha,n\rangle.
            \end{matrix}
            \right.
    \end{align}
\end{subequations}
The ratio of constants $c_p$ and $c_m$ in Eq. (\ref{eq:psi E}) is determined by imposing the constraint from Eq. (\ref{eq:eigenvalue problem b}), 
while their separate values are fixed through the normalization
\begin{equation}
    \begin{aligned}
        \langle \psi_E|\psi_E\rangle=\langle \psi_E^+|\psi_E^+\rangle+\langle \psi_E^-|\psi_E^-\rangle=1.
    \end{aligned}
\end{equation} 
The calculation, while conceptually straightforward, involves considerable algebraic manipulation. We therefore 
present only the final results as follows
\begin{subequations}
    \begin{align}
        |\psi _0 \rangle ={}&S(\xi)  |0 \rangle  |\phi _0 \rangle, \\
        |\psi_{n,\pm} \rangle ={}&S(\xi) D(\alpha_{n,\pm}) (|n-1 \rangle|\phi_1\rangle \pm |n\rangle |\phi_0\rangle) /\sqrt{2},
    \end{align}
\end{subequations}
where $S(\xi)$ and $D(\alpha_{n,\pm})$ are the squeezing and displacement
operators for the photonic field, respectively, with $\xi=\frac{1}{4}e^{-2i\phi}\ln \left( 1-\eta ^2 \right)$ and $\alpha _{n,\pm}=\mp \sqrt{n}\eta e^{-i\phi}$.
The expressions of $|\phi _0 \rangle$ and $|\phi _1 \rangle$ are given by
\begin{subequations}
    \begin{align}
        |\phi _0 \rangle ={}& c_+|g\rangle-e^{-i\phi}c_-|e\rangle,\\
        |\phi _1 \rangle ={}& c_+|e\rangle - e^{i\phi}c_-|g\rangle,
    \end{align}
\end{subequations} 
with $c_{\pm}=\left( 1\pm \sqrt{1-\eta ^2} \right)^{1/2}/\sqrt{2}$.

\section{critical QGT}

\begin{figure}[htbp]
    \centering
    \includegraphics[width=\linewidth]{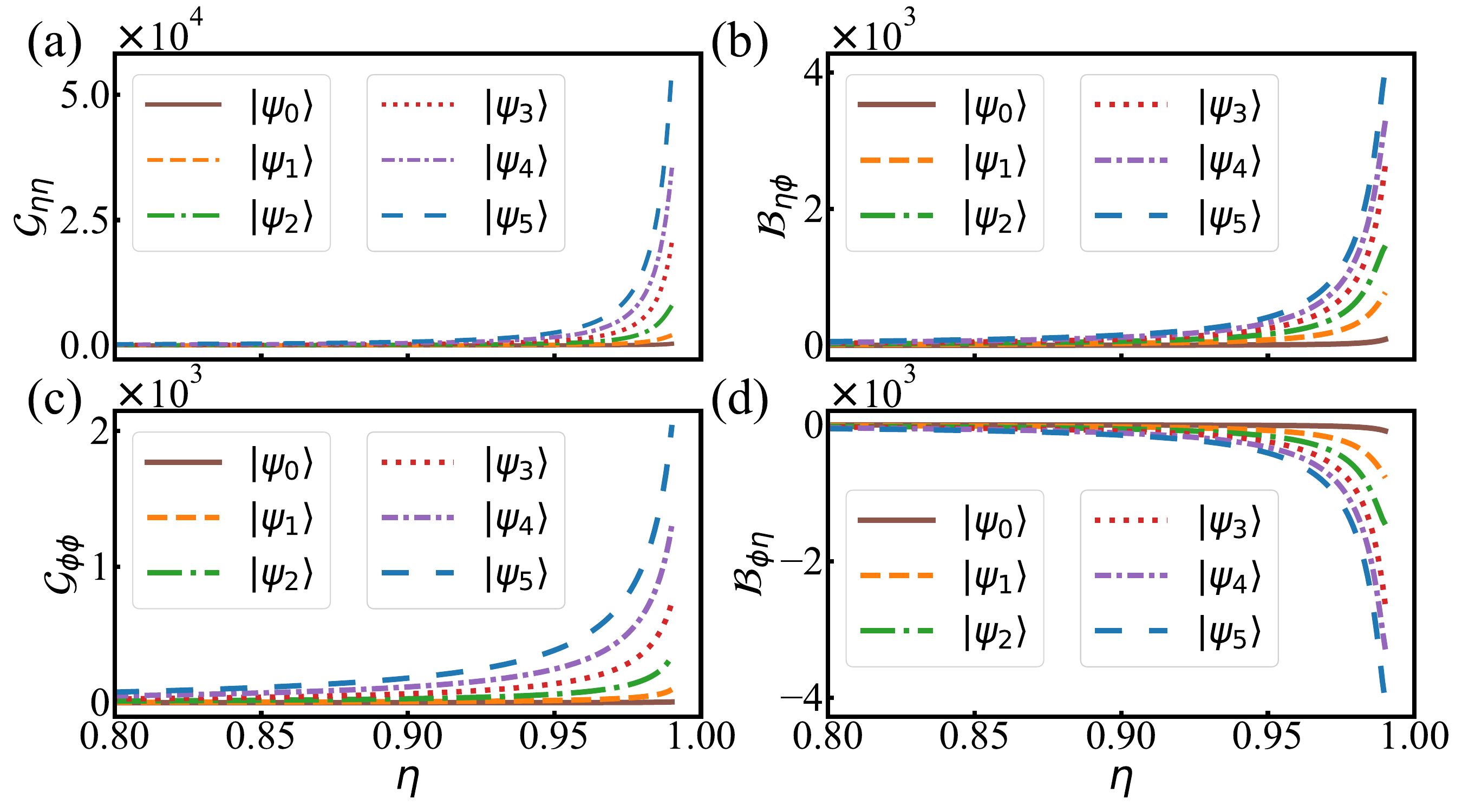}
    \caption{The quantum metric $\mathcal{G}_{\eta\eta}$ (a), $\mathcal{G}_{\phi\phi}$ (c) and the Berry curvature $\mathcal{B}_{\eta \phi}$ (b), $\mathcal{B}_{\phi \eta}$ (d)
    with respect to $\eta$ for $\phi=0$. The eigenstates $|\psi_i\rangle \ (i=0,1,2,3,4,5)$ correspond to $n = 0,1,2,3,4,5$, respectively. 
    The quantum metric $\mathcal{G}_{\eta \phi}$, $\mathcal{G}_{\phi \eta}$ and the Berry curvature $\mathcal{B}_{\eta\eta}$, $\mathcal{B}_{\phi\phi}$ are equal to zero, and not shown here.
    }
    \label{QGT all}
\end{figure}

Here, we apply the QGT to investigate the singular behavior of the driven JCM in
the vicinity of the critical points. 
We consider a nondegenerate quantum state $|\psi (\lambda) \rangle$ in a parameter space spanned by N dimensionless
parameters $\lambda(\lambda_1, \lambda_2, ..., \lambda_N)$, the QGT is defined as \cite{Provost1980Riemannian}
\begin{equation}
    \begin{aligned}
        Q_{\mu \nu}={}&\langle \partial_\mu \psi(\lambda)|[1-\mathcal{P}(\lambda)]|\partial_\nu \psi (\lambda)\rangle \label{eq:QGT1}\\
        ={}&\mathcal{G}_{\mu \nu}-\frac{i}{2}\mathcal{B}_{\mu \nu},
    \end{aligned}
\end{equation} 
where $\{|\psi (\lambda) \rangle\}$ forms a complete orthonormal
basis space and $\mathcal{P}(\lambda)=|\psi (\lambda) \rangle \langle \psi(\lambda)|$ is the Hilbert eigenspace projection operator. 
Its real component defines the quantum metric, $\mathcal{G}_{\mu \nu}={\rm Re}\{Q_{\mu \nu}\}=\mathcal{G}_{\nu \mu}$, while 
imaginary component is related to the Berry curvature, $\mathcal{B}_{\mu \nu}=-2{\rm Im}\{Q_{\mu \nu}\}=-\mathcal{B}_{\nu \mu}$.

A remarkable feature of the QGT is that its divergent behavior in the vicinity of the critical point \cite{PhysRevLett.99.095701, PhysRevLett.122.210401, PhysRevB.103.174104}. 
To elucidate this behavior, we derive a perturbative expression for QGT.
Substituting the identity operator $\mathbb{I}=\sum_m |\psi_m\rangle\langle\psi_m|$ into Eq. (\ref{eq:QGT1}) and using
\begin{equation}
    \begin{aligned}
        \langle \psi_m|\partial_\nu \psi_n \rangle=\frac{\langle \psi_m|\frac{\partial H}{\partial \nu}|\psi_n \rangle}{E_n - E_m}  \quad (m\ne n),
    \end{aligned}
\end{equation} 
which is determined by the eigenvalue equation $H|\psi_n\rangle=E_n|n\rangle$, 
we reformulate Eq. (\ref{eq:QGT1}) as follows \cite{S0217979210056335}:
\begin{equation}
    \begin{aligned}
        Q_{n,\mu \nu}={}&\sum_{m \ne n}\frac{\langle \psi_n|\frac{\partial H}{\partial \mu}|\psi_m \rangle  \langle \psi_m|\frac{\partial H}{\partial \nu}|\psi_n \rangle}{(E_m-E_n)^2}. \label{eq:QGT2}\\
    \end{aligned}
\end{equation} 
This expression confirms that the QGT components—the quantum metric tensor and Berry curvature—become singular at stationary points of the QPT, 
marked by ground-state level crossings.

We consider the QGT space formed by the parameters $\eta$ and $\phi$, as shown in Fig. \ref{QGT all}. 
Fig. \ref{QGT all}(a) and Fig. \ref{QGT all}(c) show the real parts of the QGT, i.e., $\mathcal{G}_{\eta\eta}$ and $\mathcal{G}_{\phi\phi}$, with respect to $\eta$, over the ground state $|\psi_0\rangle$ and the forefront five low-lying excited states $|\psi_i\rangle \ (i=1,2,3,4,5)$;
while Fig. \ref{QGT all}(b) and Fig. \ref{QGT all}(d) show the imaginary parts of the QGT, i.e., $\mathcal{B}_{\eta \phi}$ and $\mathcal{B}_{\phi \eta}$, as the function of $\eta$, over the same states.
It is clearly shown that, all
$\mathcal{G}_{\eta\eta}$, $\mathcal{G}_{\phi\phi}$, $\mathcal{B}_{\eta \phi}$ and $\mathcal{B}_{\phi \eta}$, in the excited states $|\psi_i\rangle \ (n=1,2,3,4,5)$, exhibit significantly greater divergence compared to the ground state $|\psi_0\rangle$. 
Furthermore, the behavior of the divergence becomes stronger as the energy level separation of $|\psi_i\rangle \ (n=1,2,3,4,5)$ from $|\psi_0\rangle$ becomes larger, 
which can also be seen from Eq. (\ref{eq:QGT2}).

% \section{Critical Bures metric}

\begin{figure}[htbp]
    \centering
    \includegraphics[width=\linewidth,page=1]{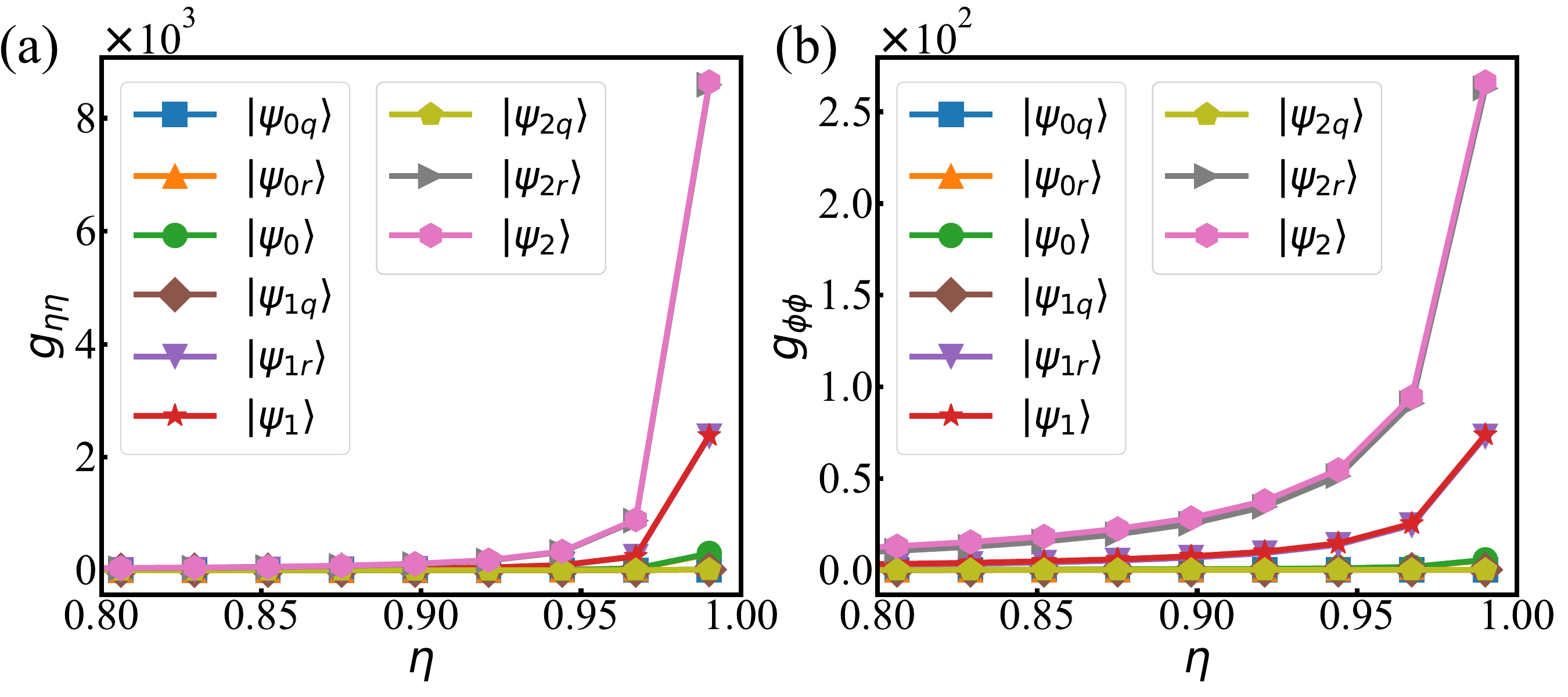}
    \caption{The Bures metric $g_{\eta\eta}$ (a) and $g_{\phi\phi}$ (b) as a function of $\eta$. 
    Here $|\psi_i\rangle$, $|\psi_{iq}\rangle$ and $|\psi_{ir}\rangle$ ($i = 0,1,2$) represent the eigenstates of the composite system, 
    the two-level subsystem and 
    the quantum field mode, respectively. 
    The $g_{\eta\phi}$ and $g_{\phi\eta}$ are equal to zero, and not shown here.
    }
    \label{Bures metric}
\end{figure}

In quantum information geometry, the Bures metric defines a natural Riemannian structure on the space of density matrices, 
which is derived from the Bures distance.
For two density matrices $\rho_1$ and $\rho_2$, 
the Bures distance is defined via the Uhlmann fidelity $\mathcal{F}\left( \rho _1,\rho _2 \right) =\left( \text{tr}\sqrt{\sqrt{\rho _1}\rho _2\sqrt{\rho _1}} \right) ^2$ \cite{UHLMANN1976273} as 
\begin{equation}
    \begin{aligned}
        dS^{2}\left( \rho _1,\rho _2 \right) =2\left( 1-\sqrt{\mathcal{F}\left( \rho _1,\rho _2 \right)} \right).
    \end{aligned}
\end{equation}
The corresponding Bures metric $g_{ij}$ is given by
\begin{equation}
    \begin{aligned}
        \sum_{i,j}{g}_{ij}(\textbf{x}) dx_i dx_j=dS^{2}\left( \rho _{\textbf{x}},\rho _{\textbf{x} + d\textbf{x}} \right),
    \end{aligned}
\end{equation} 
which quantifies the distinguishability between two neighboring density matrices, $\rho_{\mathbf{x}}$ and $\rho_{\mathbf{x} + d\mathbf{x}}$, in a parameter space $\mathbf{x}$.

As shown in Fig. \ref{Bures metric}, the Bures metric components $g_{\eta\eta}$ and $g_{\phi\phi}$ 
become divergent in the proximity of the critical point. 
The eigenstates of the composite system are denoted by  
$|\psi_i\rangle$ ($i = 0,1,2$), which describe the product state of the eigenstates of the two-level subsystem ($|\psi_{iq}\rangle$) 
and the quantum field mode ($|\psi_{ir}\rangle$), i.e., $|\psi_i\rangle=|\psi_{iq}\rangle\otimes|\psi_{ir}\rangle$. 
Moreover, it is shown that Bures metric components $g_{\eta\eta}$ and $g_{\phi\phi}$ diverge much more strongly in the excited states 
$|\psi_i\rangle (i=1,2)$ than in the ground state $|\psi_0\rangle$.
Interestingly, this divergent behavior is monotonically enhanced as the energy-level distance of $|\psi_i\rangle \  (i=1,2)$ from the ground state $|\psi_0\rangle$ increases.
It's noted that the Bures metric components $g_{\eta\eta}$ and $g_{\phi\phi}$ of the composite system 
are dominated by contributions from the quantum field mode eigenstates $|\psi_{ir}\rangle$ ($i = 0,1,2$).

\section{Experimental feasibility}
The difficulty in the experimental preparation of a target quantum state stems from the fact that, although it is theoretically well-defined, 
its direct initialization is often prohibitively challenging.
To circumvent this, a widely employed strategy leverages the quantum adiabatic theorem \cite{Lue2026Critical,RevModPhys.90.015002,farhi2000adiabatic,born_beweis_1928,101143JPSJ,PhysRevA.71.012331}. 
Such an approach initializes the system in an easily preparable eigenstate at an initial parameter value, subsequently applying a slow, 
continuous drive to evolve the Hamiltonian towards the final parameter value. 
Within our framework, the dynamics of the driven JCM can be governed by the time-dependent critical parameter $\eta =\sqrt{1-[(kt)^{4/3}+1]^{-1}}$  
\cite{PhysRevResearch.3.023062}, where $k$ determines the ramping velocity.
Initial conditions at $t=0$ are set to $|0\rangle|g\rangle$ for the ground state 
and to the superposition state $(|n-1\rangle|e\rangle \pm |n\rangle|g\rangle)/\sqrt{2}$ for the $i$th eigenstates, respectively.

However, the experimental feasibility of this approach is fundamentally constrained by the adiabatic condition, 
which requires that the driving rate ($\sim 1/t$, $t$ is the evolution time) be much smaller relative to the spectral gaps along the parameter trajectory \cite{RevModPhys.83.863,KOLODRUBETZ20171}. 
This requirement becomes particularly stringent near quantum critical regions, where the protocol's driving speed must be drastically reduced. 
Notably, the requisite slow-down is even more pronounced for the excited states.
This substantial prolongation of adiabatic evolution time introduces a fundamental limitation for implementing robust quantum information protocols under realistic experimental conditions. 
To overcome this challenge, a possible solution resorts to so-called counterdiabatic driving, which is also known as the adiabatic shortcut \cite{PhysRevLett.111.100502,pnas1619826114,pqhl_nbtk,wbbs_s8fs}. 
By introducing auxiliary control fields that actively suppress non-adiabatic excitations, it enables, in principle, a transitionless quantum evolution within an arbitrarily short duration, offering a pathway toward robust finite-time quantum state engineering in critical systems. 
Furthermore, a recent study has demonstrated an excited state preparation technique that exploits the physical coupling between photons and electrons to directly convert between ground and excited states \cite{burton2025excitedstatepreparationquantum}.

\section{Conclusions}
In conclusion, we have studied the quantum geometry associated with
the eigenstates of the driven JCM. Our results reveal that both the quantum
metric and Berry curvature exhibit critical behaviors near the
photon-blockade breakdown phase transition. The divergent features of these
quantum geometric properties for bright eigenstates are much more pronounced
than those for the dark eigenstate. Our work provides a direct insight into
critical phenomena in fully quantum-mechanical light-matter systems from the
view of differential geometry. The predicted geometric features can be
observed with present techniques in superconducting circuits.

\section{ACKNOWLEDGEMENTS}
This work was supported by the National Natural Science Foundation of China (Grants No. 12474356, No. 12475015, No. 12274080, No. 12505016),  and the Natural Science Foundation of Fujian Province (Grant No. 2025J01465). 

% \section{CONFLICT OF INTEREST}
% The authors declare no conflict of interest.

% \section{DATA AVAILABILITY STATEMENT}
% The data that support the findings of this study are available from the corresponding author upon reasonable request.

\bibliography{ref} 
\end{document}